\documentstyle[twocolumn,aps]{revtex}
\begin{document}

\draft

\input epsf

\title{Kinetic Capacity of a Protein}
\author{Robin C.\ Ball$ ^{\S}$ and Thomas M.\ A.\ Fink$ ^{\dag}$}
\address{Theory of Condensed Matter, Cavendish Laboratory,
Cambridge CB3 0HE, England \\
{\rm $^{\S}$r.c.ball@warwick.ac.uk} \quad {\rm $^{\dag}$tmf20@cam.ac.uk} \quad
{\rm $^{\dag}$http://www.tcm.phy.cam.ac.uk/$\,\tilde{}\,$tmf20/}
}

\maketitle

\begin{abstract}
The ability of a protein to recognise multiple independent
target conformations was demonstrated in \cite{FBCT00}.
Here we consider the recognition of correlated configurations,
which we apply to funnel design for a single conformation.
The maximum basin of attraction, as parametrised in our model,
depends on the number of amino acid species as $\ln A$,
independent of protein length.
We argue that the extent to which the protein energy landscape can be
manipulated is fixed, effecting a trade-off between
well breadth, well depth and well number.
This clarifies the scope and limits of protein and heteropolymer function.
\end{abstract}

\narrowtext


\input epsf

It is believed that a stable, fast folding protein requires an
energy landscape in which the native conformation is
both a deep global minimum and lies at the bottom of a basin of attraction 
sloping towards it \cite{Dill97}.
These conditions are known as thermodynamic stability and 
kinetic accessibility, respectively.
While stability may be readily achieved by suppressing the energy of the 
sequence arranged in the target conformation,
constructing a broad funnel leading towards the target has remained elusive.

The first satisfactory method of protein design, introduced by 
Shakhnovich in 1994 \cite{Shak94}, relies on the
correlation between stability and accessibility: stable sequences are 
found to fold more quickly as well.  
Minimising the energy or relative energy over sequence space 
while the conformation remains quenched to the target
\cite{Shak94,Gutin95} yields protein sequences which 
fold much more rapidly than random heteropolymers of equal length.
We have provided evidence, nonetheless, that the most stable sequences 
are not the fastest folding, and that a reduction in stability allows 
significant gain in efficiency \cite{FB97}.

In this Letter we investigate the introduction of a folding funnel above the
target conformation in the protein energy landscape \cite{FinkThesis}.
Our method of design relies on the technique of training to multiple 
targets discussed in \cite{FBCT00}.  Unlike the independent conformations 
previously considered, here our patterns are correlated to a single target 
conformation.
We find that the extent of the optimal folding funnel, as parameterised by our
model, is smaller than the conformational space and 
depends on the number of amino acid species available.
The influence of alphabet size on the folding performance of 
{\it untrained} sequences is considered in \cite{Garel89}.

Our approach to funnel design is to turn off all the monomer 
interactions (equivalent to an 
interacting system at infinite temperature) and to consider the
dynamics by which a protein would then spontaneously {\it unfold} from 
the target state into a random ensemble.
By the principle of detailed balance in equilibrium statistical mechanics,  the
ensemble of unfolding trajectories from the target state to random conformations
is equivalent to the ensemble of folding trajectories from random configurations
to the target --- but of course the former ensemble is much more easily
sampled.
Therefore, observations of unfolding will tell us how the molecule would with 
least dynamical constraint fold.

We provide estimates of the unfolding contact map based on a blob model
of unfolding.  
This is motivated by thermodynamic tractability 
and its basis in established polymer physics, despite its at times 
unrealistic representation of kinetics.
It leads to a definite proposal as to how different stages
in the unfolding contact map should be weighted in training so as to create
an optimal funnel.  

We find, however, that training to the ideal folding funnel cannot be achieved.
Remarkably, the bound on funnel size (in terms of a relaxation
length scale) is identical to the 
thermodynamic capacity derived in \cite{FBCT00}.  Taken together,
our results suggest that the extent to which the protein energy landscape 
can be manipulated --- whether it be by the introduction of multiple  
independent minima, well depth or well breadth (or a combination thereof)
--- is limited and proportional to the $\log$ of the number of 
amino acid species.


{\bf Generalisation to Weighted Training}
In a separate Letter \cite{FBCT00} we investigated the design of 
multi-stable proteins by training to a uniform superposition of contact maps.
The typical well depth of a protein of length $N$ embedded in one of the target
conformations was found to be

\begin{equation}
E_{\mu}^{\rm min} \simeq -{\textstyle \sqrt{z'\over p}} N \sigma \sqrt{\ln A}.
\label{FUNNEL_A0}
\end{equation}

\noindent where $A$ is the number of amino acid species, $\sigma$ is 
the standard deviation of the interaction energies and $z' = z-2$ is the 
effective coordination number, {\it i.e.}, the maximum number of local 
contacts excluding the backbone.
After training to a weighted superposition of contact maps, we
expect conformations associated with higher weights to have deeper wells.
The derivation of the precise dependence follows.

The total contact map is defined by summing 
over the individual maps with suitable weights,

\begin{equation}
C_{{\rm tot}_{ij}} = \sum_{\mu = 1}^p w_\mu C_{\mu_{ij}},
\label{FUNNEL_A}
\end{equation}

\noindent where $w_\mu$ is the weight associated with conformation $\Gamma_\mu$.
The minimum Hamiltonian associated with the total weighted contact map is

\begin{equation}
H_{\rm tot}^{\rm min} = {1\over 2} \sum_{ij = 1}^N C_{{\rm tot}_{ij}} 
     \tilde{U}_{ij}^* = {1\over 2} \sum_{ij=1}^N \sum_{\mu=1}^p w_\mu 
      C_{\mu_{ij}} \tilde{U}^*_{ij}.
\label{FUNNEL_B}
\end{equation}

\noindent where $\tilde{U}^*$ minimises $H_{\rm tot}$.

By analogy with calculations in \cite{FBCT00}, we re-express (\ref{FUNNEL_B})
as a sum over $H_{{\rm tot}_i}$, each minimised by the choice of $S_i$,

\begin{equation}
H_{\rm tot}^{\rm min} = \sum_{i=1}^N \min_{S_i} [H_{{\rm tot}_i}],
\label{FUNNEL_C}
\end{equation}

\noindent where $H_{{\rm tot}_i}$ is the sum over connections to monomer $i$,

\begin{equation}
H_{{\rm tot}_i} = {1\over 2} \sum_{j=1}^N \sum_{\mu=1}^p w_\mu C_{\mu_{ij}}
\tilde{U}_{ij}.
\label{FUNNEL_D}
\end{equation}

\noindent The local Hamiltonian $H_{{\rm tot}_i}$ is simply a weighted sum 
of the independent local conformational energies,

\begin{equation}
H_{{\rm tot}_i} = \sum_{\mu = 1}^p w_\mu E_{\mu_i}.
\label{FUNNEL_E}
\end{equation}

\noindent Proceeding as in \cite{FBCT00}, we approximate the distribution of
$H_{{\rm tot}_i}$ by its central limit form;
it is a gaussian with variance
$\sigma_{{\rm tot}_i}^2 = {z' \over 2} \sigma^2 \sum_{\mu=1}^p w_\mu^2.$
\label{FUNNEL_F}
This estimation is valid out to $|H_{{\rm tot}_i}|$ $\sim$
${z'\over 2} \sigma \sum_{\mu=1}^p w_\mu$.

We now consider $H_{{\rm tot}_i}$ in (\ref{FUNNEL_E}) as a sum of two terms,

\begin{equation}
H_{{\rm tot}_i} = w_\mu E_{\mu_i} + \!\!\! 
\sum_{\nu =1,\nu \neq \mu}^p \!\!\! w_\nu E_{\nu_i} 
                = H_{\mu_i} + H_{{\rm oth}_i}.
\label{FUNNEL_G}
\end{equation}

\noindent Since $H_{\mu_i}$ and $H_{{\rm oth}_i}$ are independently 
gaussianly distributed with variances

\begin{equation}
\sigma_{\mu_i}^2 = {z' \sigma^2\over 2} w_\mu^2 \quad {\rm and} \quad
\sigma_{{\rm oth}_i}^2 = {z' \sigma^2\over 2} 
\!\!\! \sum_{\nu =1,\nu \neq \mu}^p \!\!\! w_\nu^2,
\label{FUNNEL_H}
\end{equation}

\noindent the distribution of $H_{\mu_i}$ for fixed 
$H_{\mu_i} + H_{{\rm oth}_i} = H^{\rm min}_{{\rm tot}_i}$ reduces to

\begin{equation}
f(H_{\mu_i}|H^{\rm min}_{{\rm tot}_i}) \simeq c 
\exp\Big(-{\sigma_{{\rm tot}_i}^2 \over 2 \sigma_{\mu_i}^2 \sigma_{{\rm oth}_i}^2}
     (H_{\mu_i} - {\sigma_{\mu_i}^2 \over \sigma_{{\rm tot}_i}^2} 
     H^{\rm min}_{{\rm tot}_i})^2\Big),
\label{FUNNEL_J}
\end{equation}

\noindent where $c$ is a normalising constant and 
$\sigma_{{\rm tot}_i}^2 = \sigma_{\mu_i}^2 + \sigma_{{\rm oth}_i}^2$. 
The value of $H_{{\mu}_i}$ of maximum likelihood from (\ref{FUNNEL_J}) 
is given by

\begin{equation}
H_{\mu_i}^{\rm min} 
= {\sigma_{\mu_i}^2 \over \sigma_{{\rm tot}_i}^2} 
H^{\rm min}_{{\rm tot}_i},
\label{FUNNEL_AA}
\end{equation}

\noindent which reduces to 

\begin{equation}
H_{\mu_i}^{\rm min} 
= w_{\mu} E_{\mu_i}^{\rm min} 
= {w_\mu^2 \over \sum_{\mu=1}^p w_\mu^2} H^{\rm min}_{{\rm tot}_i}.
\label{FUNNEL_AA2}
\end{equation}

The minimum local Hamiltonian corresponds to the smallest of $A$ samples from
the distribution of $H^{\rm }_{{\rm tot}_i}$.
We approximate the minimum of $A$ samples
from a gaussian of zero mean and standard deviation $\sigma_{{\rm tot}_i}$ 
by \cite{FBCT00}

\begin{equation}
H_{{\rm tot}_i}^{\rm min} \simeq -\sqrt{2} \sigma_{{\rm tot}_i} \sqrt{\ln A}.
\label{FUNNEL_AA3}
\end{equation}

\noindent Substituting (\ref{FUNNEL_AA3}) into (\ref{FUNNEL_AA2}) and summing over $i$ yields

\begin{equation}
E_\mu^{\rm min} \simeq - \sqrt{z'} N \sigma \sqrt{\ln A} 
{w_\mu \over \Bigl(\sum_{\mu=1}^p w_\mu^2\Bigr)^{1\over 2}},
\label{FUNNEL_BB}
\end{equation}

\noindent This establishes how the minimised Hamiltonian distributes 
over the individual weighted configurations;
for the special case of equal weights it duly reduces to (\ref{FUNNEL_A0}).

{\bf Blob Model of Unfolding}
It is a well known trend in polymer physics that the larger scale features of
molecular conformations have systematically longer relaxation times.  
For example, for non-interacting chains with simple kink-jump dynamics, 
a subsection of $g$ monomer units has relaxation time $\tau(g)$ proportional 
to $g^2$ \cite{POLYMERS}.  
On this basis we assume that after time $t$, a spontaneously 
unfolding polymer will have equilibrated
locally up to scale $g$, such that $\tau(g)=t$, but still reflect the folded
conformation on larger scales.

This blob view of proteins, that time scales relate uniformly to 
length scales, is of course a particular and simplified outlook, 
motivated by its tractability.  
Complications which we do not address here include spatially localised 
nucleation events and specific configurational bottlenecks.
Nevertheless, it allows us to
make {\it some} quantitative predictions about the limits of the 
basin of attraction, which has long proved to be evasive.

The folded protein, which we assume to be compact and associate with
$g=1$, consists of $N$ single monomer blobs.  The contact map $C(1)$ has 
$z'$ non-zero entries in each row and column, $z'N$ non-zero entries in total.  

For the state unfolded up to length scale $g$, the protein may be thought of 
as a chain of ${N\over g}$ blobs, folded to its coarse grained original
conformation.  Accordingly, the contact map 
$C(g)$ has ${N\over g}$ intra-blob blocks along the diagonal and
${z'N\over g}$ inter-blob blocks corresponding to nearest neighbour
blobs (not along the backbone).  Scaling theories for polymer 
configurations with excluded volume would imply that the average total 
number of contacts between two neighbouring blobs be of order unity.  
Averaging over an ensemble of conformations at constant 
$g$, this requires that each of the $g^2$ entries for each blob be of order 
${1\over g^2}$.  

The total number of conformations (compact or otherwise) available to a 
protein grows as $\sim \tilde{\kappa}^N$ \cite{POLYMERS}
(not to be confused with $\kappa\simeq 1.85$ \cite{Pande94} for 
compact structures only); 
this becomes $\tilde{\kappa}^{N\over g}$ for a chain of ${N\over g}$ blobs.  
Since the product of the internal and external conformational freedoms 
of a partially relaxed protein must equal $\tilde{\kappa}^N$, a protein 
relaxed to length scale $g$ can be estimated to take on
$\tilde{\kappa}^{(N-{N\over g})}$ configurations.  It follows that the entropy 
gained in folding from a denatured configuration down to a conformation 
relaxed to length scale $g$ is 

\begin{equation}
S(g) = -k_B {N\over g} \ln\tilde{\kappa}.
\label{FUNNEL_CC}
\end{equation}

{\bf Training to a Funnel}
While an energy minimum significantly below the minimum copolymer energy 
ensures thermodynamic stability of the target
conformation, rapid convergence necessitates a funnel of kinetic pathways 
sloping towards the target.
The widest possible funnel is that which least constrains the dynamics, 
which we propose is given by the conformations sampled in 
unfolding via the blob model.  We thus consider combining the contact maps 
from different times (and values of $g$) of a noninteracting,
spontaneously unfolding compact conformation with weights $w(g)$,

\begin{equation}
C_{{\rm tot}_{ij}} = \sum_{\ln g = 1}^{\ln N} w(g) C_{ij}(g).
\label{FUNNEL_DD}
\end{equation}

\noindent The minimum Hamiltonian associated with the total contact map
then appears as

\begin{equation}
H_{\rm tot}^{\rm min} 
= {1\over 2} \sum_{ij=1}^N \sum_{\ln g=1}^{\ln N} w(g)
      C_{ij}(g) \tilde{U}^*_{ij},
\label{FUNNEL_FF}
\end{equation}

\noindent analogous to (\ref{FUNNEL_B}).
The total Hamiltonian associated with monomer $i$ is the sum of the 
individual local Hamiltonians evaluated at different values of $g$,

\begin{equation}
H_{{\rm tot}_i}^{\rm min} 
= \sum_{\ln g = 1}^{\ln N} H_i^{\rm min}(g),
\label{FUNNEL_FF2}
\end{equation}

\noindent where $H(g) = w(g) E(g)$.  
In accordance with our previous calculation, we require 
$\sigma_{{\rm tot}_i}^2$.  
We first estimate the variance in the choice of $H(g)$ available to 
a single monomer as

\begin{equation}
\sigma_{g_i}^2 \simeq {z'g\over 2} \Big({w(g)\over g^2}\Big)^2 \sigma^2,
\label{FUNNEL_GG}
\end{equation}

\noindent where ${z'g\over 2}$ is the number of contacts available to a given 
monomer 
equilibrated to scale $g$ and ${w(g)\over g^2}$ is the overall weighting 
for each one.  The variance of the local energy per monomer integrated over 
all $g$ is then

\begin{equation}
\sigma_{{\rm tot}_i}^2 \simeq \sum_{\ln g = 1}^{\ln N} \sigma^2_{g_i}
\simeq {z' \sigma^2\over 2} \int_e^N {dg\over g} g {w^2(g)\over g^4}.
\label{FUNNEL_HH}
\end{equation}

\noindent Again we wish to establish how the minimised Hamiltonian 
distributes over weighted configurations unfolded to length scale $g$.
Applying the general result (\ref{FUNNEL_AA}) yields

\begin{equation}
H_i^{\rm min}(g) 
\simeq w(g) E_i^{\rm min}(g) 
\simeq {\sigma_{g_i}^2 \over \sigma_{{\rm tot}_i}^2} 
H^{\rm min}_{{\rm tot}_i}.
\label{FUNNEL_II}
\end{equation}

\noindent Substituting (\ref{FUNNEL_AA3}) and (\ref{FUNNEL_GG}) into 
(\ref{FUNNEL_II}) and summing over $i$, the minimum 
energy associated with matching the conformation at scale $g$ can then be 
estimated as 

\begin{equation}
E^{\rm min}(g) \simeq - {z' \over \sqrt{2}} N \sigma^2 \sqrt{\ln A} {w(g) \over 
\sigma_{{\rm tot}_i} g^3}.
\label{FUNNEL_KK}
\end{equation}

In order that the training reverse the unfolding dynamics, the required funnel
must have sufficient slope, that is, $F(g) = E(g) -T S(g)< 0$.
Equating the two
expressions $T \times$ (\ref{FUNNEL_CC}) and (\ref{FUNNEL_KK}) gives

\begin{equation}
w(g) \simeq - {\sqrt{2} k_B T \ln \tilde{\kappa} \,\, \sigma_{{\rm tot}_i} \over z' \sigma^2 \sqrt{\ln A}} g^2,
\label{FUNNEL_LL}
\end{equation}

\noindent and thus $w(g) \propto g^2$.  
Unfortunately this form for $w$ is inconsistent with a convergent 
($N$-independent) evaluation of $\sigma_{{\rm tot}_i}$ in (\ref{FUNNEL_HH}).
Our assumption that the training energy could reverse the unfolding dynamics 
does not hold for all values of $g$.  

We consequently introduce the cutoff scale $g_{\rm max}$, up to which our 
funnel extends.
Substituting (\ref{FUNNEL_LL}) into (\ref{FUNNEL_HH}) and reducing the
domain of integration yields

\begin{equation}
\sigma_{{\rm tot}_i}^2 \simeq {(k_B T)^2 \ln^2 \tilde{\kappa} \over 
                z' \sigma^2  \ln A} 
                          \sigma_{{\rm tot}_i}^2 \int_e^{g_{\rm max}} dg,
\label{FUNNEL_MM}
\end{equation}

\noindent from which it follows that

\begin{equation}
g_{\rm max} \simeq {z' \sigma^2 \ln A \over (k_B T)^2 \ln^2 \tilde{\kappa} }.
\label{FUNNEL_NN}
\end{equation}

The width of our funnel, as parametrised by $g_{\rm max}$ above, increases 
strongly as folding temperature $T$ decreases.  At too low a temperature, 
however, 
the coil will collapse as a random copolymer into what we presume to be a
glassy state.  The loss in entropy resulting from collapse will 
be equivalent to $-$(\ref{FUNNEL_CC}) evaluated at $g=1$ 
(the collapsed copolymer will be fully folded).
The modest decrease in energy afforded by the minimum copolymer energy can
overcome this entropic loss only at low temperature $T_{\rm cp}$.
Equating the minimum copolymer energy $E_{\rm cp}^{\rm min}$ from
$(7)$ in \cite{FBCT00} and $T_{\rm cp}$ times the loss 
in entropy $-$(\ref{FUNNEL_CC})$|_{g=1}$ leads to

\begin{equation}
k_B T_{\rm cp} \simeq \sigma {\sqrt{z' \ln \kappa} \over \ln \tilde{\kappa}},
\label{FUNNEL_OO}
\end{equation}

\noindent and hence at $T \simeq T_{\rm cp}$,

\begin{equation}
g_{\rm max} \simeq {\ln A \over \ln\kappa},
\label{FUNNEL_PP}
\end{equation}

\noindent which is identical to the form of $p_{\rm max}$ derived in
\cite{FBCT00}.

\begin{figure}[h!]
  \begin{center}
    \leavevmode
    \epsfxsize=6.85cm
    \epsfbox{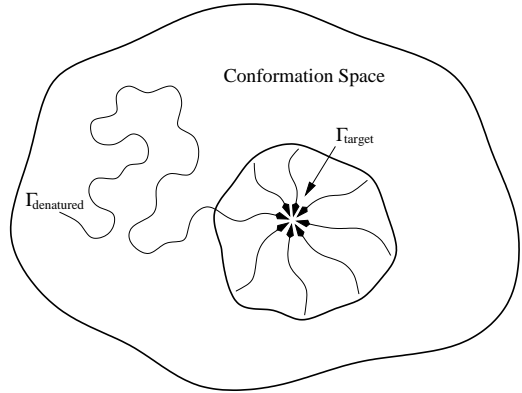}
    \end{center}
    \caption[Folding in the presence of a funnel.]
    {Folding in the presence of a funnel.  The denatured protein wanders through
    conformation space until it matches the target structure coarse-grained to
    length scale $g_{\rm max}$, after which the funnel quickly guides the 
    protein towards the target.}
  \label{blob_funnel}
\end{figure}

\begin{figure}[!t]
  \begin{center}
    \leavevmode
    \epsfxsize=6.7cm
    \epsfbox{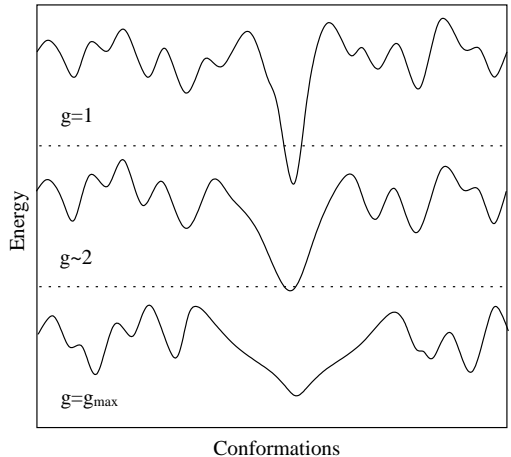}
    \vspace{0.15in}
    \caption[Energy landscapes of sequences trained to have increasingly
    broad funnels.]
    {Energy landscapes of sequences trained to have increasingly broad funnels.
    Maximising stability (top) corresponds to a deep, narrow well.
    As the length scale $g$ to which the funnel extends increases,
    the depth of the target well is reduced; at $g = g_{\rm max}$, 
    the slope of the funnel is no longer sufficient to provide a
    free energy minimum (bottom).
    }
  \label{funnelcapacity_Eland}
  \end{center}
\end{figure}

{\bf Discussion of Capacity} 
That the bound on the folding funnel $g_{\rm max}$ is less than $N$ implies
the extent of the achievable folding funnel is less than the 
conformational space of the protein.
Folding at finite temperature cannot be made as direct as unfolding at 
infinite temperature.
The cutoff $g_{\rm max}$ is the length scale 
of the structure below which the energy
landscape corresponding to the trained sequence is characterised by a funnel.
Above $g_{\rm max}$, the protein must organise itself into the desired (coarse 
grained) conformation without the help of kinetic guidance, that is, it must
traverse an effective copolymer landscape (Figure \ref{blob_funnel}).
What happens to the protein energy landscape upon increasing the width of 
the funnel?
As $g \rightarrow g_{\rm max}$, the slope of the funnel becomes sufficiently 
shallow such that, at $g = g_{\rm max}$, the decrease in energy no longer 
overcomes the loss of entropy (Figure \ref{funnelcapacity_Eland}); the well
ceases to be a free energy minimum.

Consider the protein as a sequence of $N/g_{\rm max}$ blobs, 
each of size $g_{\rm max}$.  The benefit of 
the funnel is realised once the chain of blobs folds to its coarse grained
target state.
Assuming this statistical bottleneck to be the rate determining step,
the time necessary for the protein to fold is reduced by the factor 
$\kappa^{-(1 - 1/g_{\rm max})N}$, which is significant even for small values of $g_{\rm max}$.


{\bf Manipulation of the Energy Landscape}
In both the thermodynamic \cite{FBCT00} and kinetic contexts, the extent 
to which the protein energy landscape can be manipulated is limited by 
${\ln A \over \ln \kappa}$, where $A$ is the number of amino acid
species and $\kappa$ is the compact conformational freedom per monomer.
Like squeezing one end of a balloon at the expense of inflating the other,
further deformation of the energy landscape is counter-balanced by its
relaxation elsewhere.

The agreement between the bounds on protein memory, on the one hand, and the 
basin of attraction, on the other, was unexpected.  
Taken together, these results suggest that the engineering of proteins 
and heteropolymers is constrained by a fixed budget.  
The finite freedom in the sequence can be invested in various attributes: 
in well number, well breadth and well depth.
A reduction in expenditure in one allows increased investment in another.  

In particular, our results suggest that
thermodynamic stability and kinetic accessibility, while correlated
over a significant region, are in conflict near the extremes of either;
maximally stable sequences are not the fastest folding and the fastest
folders are not the most stable.  (We presented preliminary evidence to 
this end in \cite{FB97}).  
Accordingly, thermodynamically oriented sequence
design need not select for the fastest folding proteins and a reduction 
in stability admits increased accessibility.
If Nature has designed proteins to fold as quickly as possible, we would
expect only marginal stability in the native conformation.  
The preceding premise might be established by observation of normal and 
mutated naturally occurring proteins.

Notably, the bound on manipulating the energy landscape is independent 
of protein length; the diversity of protein function grows with 
alphabet size only.
The large (relative to $\kappa$) amino acid alphabet found in
Nature is crucial to the variety of protein function within the cell
or in multicellular organisms.
To the extent that heteropolymer models are intended to provide 
insight into proteins, their alphabet sizes should reflect this.
Elementary representations, such as frequently studied H-P models, 
are not able to effect the thermodynamic and kinetic diversity possible with 
larger alphabets.

Perhaps most interesting is the increased scope for protein and heteropolymer
function.
The discovery that prions fold to multiple conformations \cite{SP98}
has extended our notion of heteropolymer behaviour beyond familiar
protein collapse.
We have presented arguments that the energy landscape may, within limits,
be tailored to effect function heretofore unobserved.
Further discovery of novel protein mechanisms should prove fascinating.


\end{document}